\documentclass[prl,aps,amssymb,twocolumn,superscriptaddress]{revtex4}
\usepackage{psfig}

\def\braket#1#2{\langle#1\vert#2\rangle}
\def\br{{\bf r}}
\def\bk{{\bf k}}

\def\pmb#1{\setbox0=\hbox{#1}%
 \hbox{\kern-.025em\copy0\kern-\wd0
 \kern.05em\copy0\kern-\wd0
 \kern-0.025em\raise.0433em\box0} }

\begin{document}

\title{
Ballistic conductance of Ni nanowire with a magnetization reversal
}

\author{A. Smogunov}
\affiliation{SISSA, Via Beirut 2/4, 34014 Trieste (Italy)}
\affiliation{INFM, Democritos Unit\`a di Trieste Via Beirut 2/4, 34014
Trieste (Italy)}
\author{A. Dal Corso}
\affiliation{SISSA, Via Beirut 2/4, 34014 Trieste (Italy)}
\affiliation{INFM, Democritos Unit\`a di Trieste Via Beirut 2/4, 34014
Trieste (Italy)}
\author{E. Tosatti}
\affiliation{SISSA, Via Beirut 2/4, 34014 Trieste (Italy)}
\affiliation{INFM, Democritos Unit\`a di Trieste Via Beirut 2/4, 34014
Trieste (Italy)}
\affiliation{ICTP, Strada Costiera 11, 34014 Trieste(Italy)}

\date{\today}

\begin{abstract} 
The approach proposed by Choi and Ihm for calculating the ballistic conductance of
open quantum systems is generalized to deal with magnetic transition metals.
The method has been implemented with ultrasoft pseudopotentials and plane wave basis set 
in a DFT-LSDA ab-initio scheme.
We present the quantum-mechanical conductance calculations for monatomic 
Ni nanowire with a single spin reversal. We find that a spin reversal 
blocks the conductance of $d$ electrons at the Fermi energy of the Ni 
nanowire. On the other hand, two $s$ electrons (one per each spin) 
are perfectly transmitted in the whole energy window giving $2G_0$ 
for the total conductance. The relevance of these results in connection
with recent experimental data is discussed.
\end{abstract}
\maketitle

Understanding electron conduction through atomic and molecular wires connecting two
macroscopic electrodes is presently a very active research topic. Conductance
measurements for metal nanocontacts display flat plateaus and abrupt drops during
the elongation attributed to atomic rearrangements. The last conductance step 
before breaking, corresponding to a monatomic nanowire,  
has a value $2G_0$ ($G_0=e^2/h$ is the conductance quantum per spin)
in monovalent metals such as gold \cite{ck} 
which can be rationalized with the free propagation of 2 (1 per each spin) valence $s$-electrons.  
Surprisingly, a low value of the last conductance step was also 
observed in nanowires 
and nanocontacts of
transition metals having partially occupied $s$ and $d$ states.
Oshima et al. \cite{oshima}, who worked in vacuum, at variable temperature, 
and with the possibility of a magnetic field, found a minimal conductance 
step in Ni nanocontact preferentially near 2 and 4 at RT and zero field, 
near 4 at 770 K and zero field, and near 3 (occasionally near 1) at RT 
with an applied
magnetic field. 
Ono et al. \cite{ono}, reported again 2 for Ni in zero field, and 1 
for Ni in a field. Recently, Ugarte et al. \cite{ugarte} observed 1 conductance quantum 
at room temperature and zero magnetic field in atomic chains (the ultimate configuration before 
the nanocontact rupture) of transition metals such as Co and Ni. 

Several theoretical methods are available to study the transport
properties of atomic scale conductors. Among these, there are methods 
based on nonequilibrium Green's functions combined with a localized
basis set \cite{taylor,palacios}. A method based
on a system-independent wavelet basis set to calculate the 
Green's function \cite{wavelet} has been also presented recently.  
Another approach for conductance calculation relies on the direct solution 
of the scattering problem
for an open system and on the application of the Landauer-Buttiker 
formula relating the 
ballistic conductance to the total transmission coefficient at the Fermi energy. 
The scattering approach has been applied by Lang and co-workers \cite{lang},
and Tsukada and co-workers \cite{hirose} using the jellium model for
the electrodes. The layer KKR approach has been used to study the 
spin-dependent tunneling in magnetic tunnel junctions \cite{lkkr}. 
Lately Wortmann and co-workers \cite{wortmann} formulated a transfer
matrix approach based on computation of single-electron 
Green function in the linearized augmented plane wave (LAPW) basis. 

We are interested in a formulation based on plane waves and ultrasoft pseudopotentials,
a calculational technology which we are applying to other problems
involving transition metals \cite{dalcorso1}.
Recently, Choi and Ihm \cite{choi} proposed a first-principles method based on plane waves to solve 
the scattering problem with real atomic contacts. In their formulation atoms are described by the 
Kleinman-Bylander-type pseudopotentials. We generalized this scheme to deal with ultrasoft pseudopotentials 
\cite{vanderbilt} in order to apply it to magnetic transition metals.

In a recent paper \cite{our1} we showed that a magnetization reversal (sharp domain wall) built inside 
a Ni nanowire effectively blocks the $d$ channels allowing only two $s$ electrons (one per each spin) to propagate 
and thus leaving only two conductance channels. That result was argued for an infinite monatomic Ni wire with 
a periodically repeated spin reversal by simply counting the number of bands crossing the Fermi level.   
However, no actual quantum-mechanical calculation of conductance was carried out. Now we substantiate our 
earlier suggestion by calculating the ballistic conductance of monatomic Ni nanowire
with a single spin reversal. We present here the intrinsic conductance of monatomic nanowires
due to only the scattering on the magnetization reversal and do not consider any additional contact 
resistance arising from the tip-wire junctions.   

\section{Method}
We study electron transport in the linear regime (in vanishingly small applied bias)
in the open quantum system consisting of a scattering region ($0<z<L$) attached to the left 
($z<0$) and to the right ($z>L$) to semi-infinite generic electrodes. We assume that the 
electrons move ballistically in the self-consistent potential and all elastic scattering 
takes place in the scattering region which can be a monatomic wire, a molecule or just 
a single atom. The self-consistent potential is obtained performing ground state DFT 
calculations with a supercell containing the scattering region and sufficiently large pieces 
of left and right electrodes to get the bulk-like potential for the electrodes. 
A wave propagating at energy $E$ with $\bk_\perp$ in the $xy$ plane   
incident from the left electrode on the scattering region will form 
a scattering state $\Psi_{{\bf k}_\perp}$ which due to the supercell geometry has 
a Bloch form in the $xy$ direction: 
\begin{equation} 
\Psi_{{\bf k}_\perp}(\br_{\perp}+{\bf R}_\perp,z;\sigma)=e^{i{\bf k}_\perp 
{\bf R}_\perp} \Psi_{{\bf k}_\perp}(\br_\perp,z;\sigma),
\label{bloch} 
\end{equation} 
where ${\bf R}_\perp$ are wire supercell lattice vectors.
The spin-polarized systems are treated within the local spin-density approximation (LSDA). 
Therefore electrons with different spin 
directions move independently in different self-consistent potentials and the scattering problem 
can be solved separately for the two spin polarizations. In the same way, different ${\bf k}_\perp$ do not mix 
and can be considered separately. We will therefore omit both ${{\bf k}_\perp}$ and $\sigma$ from 
now on. Note that if the supercell is sufficiently large in the $xy$ plane 
one can limit the calculation to the two-dimensional $\Gamma$ point, $\bk_\perp=0$.

We describe the atoms of transition metals such as Ni by ultrasoft pseudopotentials \cite{vanderbilt}.
In the ultrasoft pseudopotential scheme the stationary state $\Psi$ at the energy $E$ 
is a solution of single-particle Kohn-Sham equation which as in Ref.\cite{our2} we write in the form:
\begin{equation} 
\left[-\nabla^2+V_{\rm eff}(\br)\right]\Psi(\br)+ 
\sum_{Imn}\tilde D^I_{mn}\braket{\beta^I_n} 
{\Psi}\beta^I_m(\br-{\bf R}_I)=E\Psi(\br). 
\label{uppequation} 
\end{equation} 
with $\tilde D^I_{mn}=D^I_{mn}-Eq^I_{mn}$. The coefficients $D^I_{mn}$ and $q^I_{mn}$ characterize 
the pseudopotential which is constructed using the set of projector functions $\beta^I_m$ associated
with atom $I$ (see Refs.\cite{vanderbilt,laasonen}).

Deep within the electrodes the scattering state $\Psi$ originating from
the rightward propagating Bloch wave $\psi_j$ in the left electrode 
has an asymptotic form:
\begin{displaymath} 
\Psi= \left\{
\begin{array}{ll} 
\psi_j+\sum\limits_{i\in L}r_{ij} 
\psi_{i} 
, & \quad z<0 
\\ 
\quad\sum\limits_{i\in R}t_{ij} 
\psi_{i} 
, & \quad z>L 
\end{array} 
\right. 
\end{displaymath} 
where summation over $i\in L~(i\in R)$ includes the generalized Bloch states 
in the left (right) electrode at energy $E$ which propagate or decay to the left 
(right). The generalized Bloch states (both propagating and decaying)
constitute the so-called complex band structure of a solid. In Ref.\cite{our2} 
we presented in detail the calculation of the complex bands with ultrasoft
pseudopotentials. Having obtained the complex bands for the electrodes we calculate the transmission 
and reflection coefficients $t_{ij}$ and $r_{ij}$ by generalizing in a straightforward way 
the approach by Choi and Ihm \cite{choi} to the case of ultrasoft pseudopotentials (the details
will be presented elsewhere).
 
The ballistic conductance $G$ is related to the total transmission $T$ at the Fermi energy by the 
Landauer-B${\rm \ddot{u}}$ttiker formula $G=G_0T$ ($G_0=e^2/h$ is the conductance
quantum per spin). The total transmission is given by:
\begin{equation}
T=\sum_{ij}|T_{ij}|^2={\rm Tr}[{\bf T}^+{\bf T}],
\label{tran}
\end{equation}
where {\bf T} is the matrix of normalized transmission amplitudes
$T_{ij}=\sqrt{I_i/I_j}\cdot t_{ij}$ and $I_j$ is the probability current of the
state $\psi_j$ in the $z$ direction. Note that only rightward propagating states
in both left and right electrodes should be considered so the matrix {\bf T}
is of dimensions $M_RM_L$ where $M_L$ and $M_R$ are the number 
of propagating modes in the left and right electrodes, respectively.

The eigenvectors of the Hermitian matrix ${\bf T}^+{\bf T}$ determine the coefficients 
of a unitary transformation from the set of Bloch states $\psi_j$
to the nonmixing conductance eigenchannels \cite{channels}. 
In the eigenchannel basis the matrix ${\bf T}^+{\bf T}$ is diagonal. 
Calling its eigenvalues $T_i$, the conductance is a sum of independent 
contributions from each eigenchannel:
\begin{equation}
G=G_0\sum_i T_i,
\end{equation} 
where $T_i$ gives the transmission probability for $i$-th eigenchannel.

In order to calculate the total transmission $T$ one needs to know the
current $I_j$ carried by the propagating Bloch state $\psi_j$ in the $z$ direction. 
The $\psi_j$ is actually pseudo-wave function which coincides with 
the all-electron wave function only outside the core regions. The true current does not
coincide with the pseudo-current, and their difference is not trivial. We recently
showed how the pseudo-current should be augmented to obtain the real current \cite{our3}
and that method is implemented in the present calculation. 

The calculation of ballistic conductance then proceeds as follows. First, we perform
the supercell DFT electronic structure calculation with the plane-wave code 
({\tt PWscf}) \cite{pw} to get the self-consistent potential $V_{\rm eff}$ and the 
screened coefficients $D^I_{mn}$. For
spin-polarized systems both $V_{\rm eff}$ and $D^I_{mn}$ will depend on the 
spin direction. Second, we calculate the complex band structures of the left and right
electrodes. The unit cells of the electrodes are chosen within the supercell but
far enough from the scattering region where the potential becomes bulk-like
(for details concerning complex band structure calculations see Ref.\cite{our2}). 
At last we calculate as in Choi and Ihm's work the transmission coefficients $t_{ij}$ 
for each rightward propagating state $\psi_j$ in the left electrode and obtain the total 
transmission $T$ and eigenchannel probabilities
$T_i$ by diagonalizing ${\bf T}^+{\bf T}$ as described above. 
  
\section{Results}
We apply the scheme outlined in the previous section to calculate the ballistic conductance
of monatomic Ni nanowire with a single spin reversal. No contact scattering resistance 
is considered here so that the left and right electrodes are merely monatomic wires magnetized in
opposite directions. We consider a wire the left (right) half of which has a positive 
(negative) magnetization. A supercell containing 12 atoms (6 with up and 6 with down magnetization)
was used to calculate the self-consistent potential and the total magnetization was constrained to be zero
(for computational details see Ref.\cite{our1}). 
The scattering region and the unit cell of the left and right part of the wire are shown schematically
in Fig.1. Symmetry requires that the total transmission be the same for electrons of both spin directions.
Therefore we chose to consider only electrons with down spin polarization incident from the left on the
spin reversal region. The complex bands of the left (right) electrodes needed for the transmission calculation
are those shown on the right (left) panel in Fig. 2 of Ref.\cite{our2}.  

\begin{figure}
\hspace*{-5mm}
\psfig{file=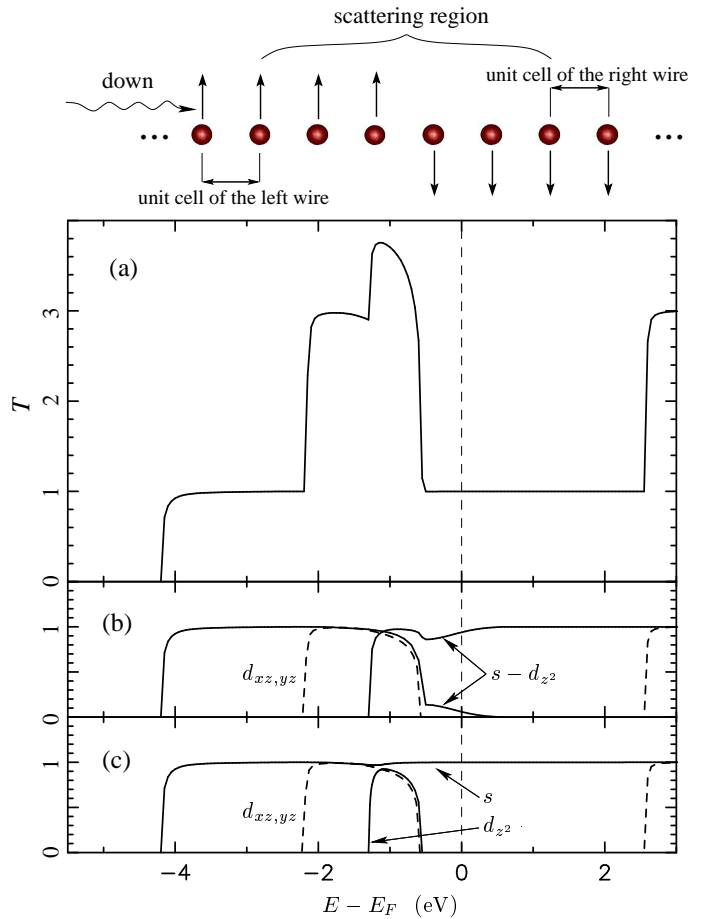,width=9.3cm,angle=0}
\caption[]
{
The electron propagation through a spin reversal in a monatomic Ni nanowire:
a) total transmission; b) contributions from each band; 
c) transmission eigenvalues $T_i$.
The bands and conducting channels are labeled by their main atomic character.
The Fermi energy is chosen to be zero (dashed vertical lines).
}
\label{fig1}
\end{figure}

In Fig. 1a and 1b we present the total transmission and the contributions from each band,
respectively, as a function of energy for the Ni nanowire with a spin reversal.
A propagating state in the left part of wire can propagate
through the spin reversal and contribute to the total transmission only if it can be matched with the 
appropriate propagating state to the right. For example, at the Fermi energy there are 6 propagating
states in the left part of the wire. However, only two $s-d_{z^2}$  states have nonzero transmission 
(Fig. 1b) since in the right piece of the wire there is only one $s-d_{z^2}$ propagating state. The $d$ states 
can be matched only with decaying states in the right part of the wire and are therefore completely reflected
by the spin reversal. At an energy of about $E=E_F-0.6$ eV the two $d_{xz,yz}$ states start contributing
to the total transmission because there are two states of the same symmetry in the right part of wire.  
Note that the $d_{xy,x^2-y^2}$ states can not get through the spin reversal at any energy
and never contribute to conductance.

Fig. 1c shows eigenchannel transmissions $T_i$. 
The two $s-d_{z^2}$ states are combined to form a single perfectly propagating conductance channel (mostly $s$ like). 
Another channel (mostly $d_{z^2}$) orthogonal to the first 
appears at energy about 0.6 eV below the Fermi energy.  It is quite narrow and has a smaller transmission
probability (about 0.9 at the maximum). 
The two $d_{xz,yz}$ states form two eigenchannels of the same symmetry. They contribute
to the total transmission between 0.6~eV and 2.2~eV below the Fermi energy and have almost unit
transmission at the maximum.

The total ballistic conductance for a wire subject to left and right chemical potentials $\mu_L$ and $\mu_R$ 
($\mu_L>\mu_R$) is:
\begin{equation}
G={G_0\over \mu_L-\mu_R}\sum_{\sigma}\int_{\mu_R}^{\mu_L} T_{\sigma}(E)dE.
\end{equation}
For $\mu_L\to\mu_R$, when $G=G_0\sum_{\sigma}T_{\sigma}(E_F)$, 
we obtain an ultimate conductance of $2G_0$ for a monatomic Ni nanowire that disagrees with the experimental
observation of a value rather closer to $G_0$ \cite{ugarte}. We should stress however that the exact equivalence
of $s$ channels for the two spin polarization, leading to the factor 2 in our result, is likely to be
an artifact of our chosen symmetrical geometry. In reality, there will generally be no up-down spin symmetry, 
and there might also 
be a more complex magnetic configuration. It seems possible that in such more complex geometries one of the $s$
spin channels could be blocked as well, which would lead to $G_0$ as observed. We plan to elaborate further 
on this line in later work.   
 
\section{Conclusions}
We generalize to magnetic transition metals the approach of Choi and Ihm for calculating the ballistic conductance
of an open quantum system. The method has been implemented with ultrasoft pseudopotentials and plane wave basis set 
in a DFT-LSDA ab-initio scheme. We applied our method to calculate the ballistic conductance of monatomic Ni wire
with a single spin reversal. We found that at the Fermi energy there is only one conductance channel per spin ($s$ like)
formed by two $s-d_{z^2}$ propagating states while all $d$ electrons are completely reflected by a spin reversal 
as previously guessed from band structure calculations \cite{our1}. 
The $s$-like channel has in fact a unit transmission probability at any energy. 

\section{Acknowledgments}

This work was sponsored by MIUR, COFIN01 and COFIN03, FIRB RBAU\O1LX5H, 
INFM (sections F, G, ``Iniziativa Trasversale calcolo parallelo''),
EU Contract ERBFMRXCT970155 (FULPROP). 
Calculations were performed on the IBM-SP4 at CINECA,
Casalecchio (Bologna).

\end{document}